\documentstyle[preprint, aps]{revtex}
\tightenlines
\begin{document}

\draft

\title{
Preparation and Superconductivity of a $MgB_2$ Superconducting Tape
}

\author{
G.C. Che\footnote{Corresponding author:  gcche@ssc.iphy.ac.cn }, S.L. Li, Z.A. Ren, L. Li, S.L. Jia, Y.M. Ni, H. Chen, C. Dong, J.Q. Li, H.H. Wen, Z.X. Zhao
}
\address{
National Laboratory for Superconductivity,
Institute of Physics and Center for Condensed Matter Physics,
Chinese Academy of Sciences, P.O. Box 603, Beijing 100080, China\\
}

\maketitle

\begin{abstract}
A $MgB_2$ superconducting tape has been successfully prepared by a new way of boron diffusion into Mg tape. This tape superconductor has a $T_c$ of 39 K with a sharp transition width of 0.4 K, high bulk critical current density $j_c$ of $6.7\times 10^5 A/cm^2$ in zero field at 6 K, higher irreversibility line comparing to that of a high pressure synthesized bulk sample and very small crystalline grain with size in the scale of 10-200 nm.
\end{abstract}


%%%%%%TEXT%%%%%%

The recent discovery of superconductivity with T$_c$ $\approx$ 40 K\cite{nagamatsu} in the intermetallic compound $MgB_2$ has stimulated intense investigations both on fundamental mechanism for the superconductivity and practical applications. On the practical application, Canfield {\it et al.}\cite{canfield} and Cunningham {\it et al.}\cite{cunningham} have reported a method to convert commercially available boron fibers into superconducting $MgB_2$ wire. Takano {\it et al.}\cite{takano} and Ren {\it et al.}\cite{ren} have measured the critical current density $J_c$ of hot-pressed pellets of the material. Superconducting epitaxial $MgB_2$ thin films have been prepared by Kang {\it et al.}\cite{kang}, Blank {\it et al.}\cite{blank} and Sbinde {\it et al.}\cite{sbinde} {\it et al}. The studies for the tape materials and films are important to the industrial applications of $MgB_2$ superconductors. In this paper, we report on the preparation and superconductivity of a tape $MgB_2$.

Tape $MgB_2$ was prepared by diffusing boron into a Mg tape. Starting materials were bright Mg tape of 0.3 mm thickness and fine boron powder. The starting Mg tape was buried in the boron powder and a pressure is applied uniformly on the direction perpendicular to the tape. The resulted pellets were wrapped with Ta foil and in turn enclosed in an evacuated quartz tube. The quartz tube was placed in a box turnace and heated to 850 $^{\circ}C$ at a rate of 150 $^{\circ}C$/h and kept at this temperature for 0.6 h, followed by furnace-cooling to room temperature. It should be noted that calcining temperature and short calcining time are the key factors to abtain the tape $MgB_2$. Tape $MgB_2$ cannot be obtained when calcining temperature is higher than 900 $^{\circ}C$ and calcining time is longer than 0.6 h due to the extreme diffusion of Mg. When calcining temperature is lower than 800 $^{\circ}C$ and calcining time is less than 0.6h, the boron powder can not fully diffuse into the Mg tape and the obtained tape $MgB_2$ superconductors have a wide superconducting transition.

X-ray diffraction (XRD) analysis was performed on an M18X-AHF type diffractometer with $CuK_{\alpha}$-radiation. PowderX program was used for lattice parameter calculations. Resistance curves were measured by the four-probe technique, while magnetization were measured by a superconducting quantum interference device ( SQUID, Quantum Design MPMS 5.5 T ) and a vibrating sample magnetometer ( VSM, Oxford 3001 ) respectively.

Figure 1 shows the XRD pattern of tape $MgB_2$. In the XRD pattern, MgO diffraction peaks of (200) and (220) in 2$\theta$ $\approx$ 42$^{\circ}$ and $\approx$ 62$^{\circ}$, which can normally be observable in bulk samples made by other techniques, can not be detected here. This indicates that our method effectively prevents the oxydization of the Mg tage in the reaction process and the sample is almost monophase. The XRD pattern indicates further that the crystalline grains in the sample is not orientated. The calculated lattice parameters are a = 0.3084(5) nm and c = 0.3525(5) nm.

Figure 2(b) shows the zero-field-cooled(ZFC) and field-cooled(FC) DC magnetization curves of the tape $MgB_2$. The ZFC curve was obtained by zero-field cooling the sample to 4 K, then applying an external field of 10 Oe, and then warming up to 300 K. FC curve is measured by applying an external field of 10 Oe at a temperature above $T_c$, then cooling the sample to 4K and subsequently warming up to 300K. Figure 2(b) shows that the sample has a superconducting transition at 37.5 K. The temperature dependence of the resistance of the $MgB_2$ is shown in Figure 2(a). It is clear that the sample has an superconducting onset transition at 39 K, with transition width smaller than 0.5 K.

Figure 3 shows the results of TEM observation. Figure 3(a) shows the bright-field TEM image, which reveals that the sample has small crystalline grain size, about 10-200 nm, being much smaller than 0.4-8 $\mu m$ for the $MgB_2$ prepared by high pressure synthesis\cite{li}. The electron diffraction (Figure 3(b)) of the sample indicates that the crystalline grains is not orientated, which is in agreement with the result obtained by X-ray diffraction. The small crystalline grain size of our sample might be attributed to the pecularity of this preparation method and short calcining time.

The magnetization hysteresis loops (MHL) of the tape $MgB_2$ measured at several temperatures 6, 14, 22, 30, 35 K up to a magnetic field of 8 T are presented in Figure 4(a). It should be noted here that, just like in other bulk samples, giant flux jumps will appear when the temperature is below 6 K. The bulk critical current density $j_c$ has been calculated according to the Been critical state model via $j_c = 20 \Delta M/Va(1-a/3b)$ and shown in Figure 4(b). It is clear that the $j_c$ in zero field is $\sim$ $6.7\times 10^5 A/cm^2$ at 6 K. This value is among the highest ones ever reported for bulk samples. To estimate the strength of quenched disorder of tape $MgB_2$, we can roughly calculate the parameter $j_c/j_0$, where $j_0$ is the depairing current density. Finnemore et al.\cite{finnemore} have given that the coherence distance $\xi _0$ = 5.2 nm and the penetration depth $\lambda _0$ = 140 nm. By using the expression $j_0 = cH_c/3\sqrt{6} \pi \lambda$ and $H_{c}(0) = \Phi _0/2\sqrt{2}\pi\lambda_0\xi_0$\cite{blatter}, we get $j_0 \approx 9.28\times 10^7 A / cm^2$ and $j_{c0} / j_0> 0.7\%$. Compared with $j_c/j_0 <0.1\%$ in high temperature superconductors (HTSC), the pinning strength of tape $MgB_2$ is strong. The performance of this tape in a high magnetic field is even more obvious. For example, the critixcal current density at 6 K and 1 T is about $ 4.18 \times 10 ^5 A / cm ^2$. This is about 3 times higher than the high pressure synthesized bulk sample\cite{ren,blank,wen}. This strong pinning in present tape may be attributed to the large surface area of crystalline grains and its density. In Fig.5 the irreversibility line has been calculated by taking a criterion of $J_c = 100 A / cm^2 $ and presented together with that of a high pressure systhesized bulk sample. It is clear that both the critical current density and the irrversibility line of the present tape are higher than that of the high pressure synthesized bulks.

In conclusion, a tape $MgB_2$ superconductor has been successfully prepared by a new way of boron diffusion into Mg tape. This tape $MgB_2$ superconductor has a $T_c$ of 39 K with a sharp transition width of $\sim$ 0.4 K, high $j_c$ of $6.7\times 10^5 A/cm^2$ in zero field at 6 K and very small crystalline grain size of 10-200nm.\\

ACKNOWLEDGEMENTS:\\
This work is supported by the Ministry of Science and Technology of China ( NKBRSF-G 19990646 ).

\begin{figure}
\caption{XRD pattern of tape $MgB_2$.}
\end{figure}

\begin{figure}
\caption{Temperature dependence of resistance (a), and ZFC and FC magnetization curves (b) of tape $MgB_2$}
\end{figure}

\begin{figure}
\caption{Bright-field TEM image (a) and electron diffraction pattern of tape $MgB_2$ (b).}
\end{figure}

\begin{figure}
\caption{Field dependence of the magnetization (a) and the critical current density (b) calculated from the Been critical state model for tape $MgB_2$.}
\end{figure}

\begin{figure}
\caption{The irreversibility line of the present tape and the high pressure synthesized bulk sample.}
\end{figure}


\begin{references}
\bibitem{nagamatsu} J. Nagamatsu, N. Nakagawa, T. Muranaka, Y. Zenitani and J. Akimitsu, Nature {\bf 410}, 63(2001).
\bibitem{canfield} P. C. Canfield, D. K. Finnemore, S. L. Budko {\it et al.}, cond-mat/0102289.
\bibitem{cunningham} C. E. Cunningham, C. Petrovic, G. Lapertot {\it et al.}, cond-mat/0103390.
\bibitem{takano} Y. Takano, H. Takeya, H. Fujin {\it et al.}, cond-mat/0102167.
\bibitem{ren} Z. A. Ren, G. C. Che, Z. X. Zhao, H. Chen, C. Dong, Y. M. Ni, S. L. Jia, H. H. Wen, Chin. Phys. Lett. {\bf 18}, 589(2001).
\bibitem{kang} W. N. Kang, H. J. Kin, E. M. Choi {\it et al.}, cond-mat/0103179.
\bibitem{blank} D. H. A. Blank, H. Hilgenkamp, H. Brinkman {\it et al.}, cond-mat/0103543.
\bibitem{sbinde} S. R. Sbinde, S. B. Ogale, R. L. Greene {\it et al.}, cond-mat/0103542.
\bibitem{li} J. Q. Li, L. Li, Y. Q. Zhou {\it et al.}, Chin. Phys. Lett. {\bf 18}, 680(2001).
\bibitem{finnemore} D. K. Finnemore {\it et al.}, cond-mat/0102114.
\bibitem{blatter} G. Blatter {\it et al.}, Rev. Mod. Phys. {\bf 66}, 1125(1994)
\bibitem{wen} H. H. Wen, S. L. Li, Z. W. Zhao, Y. M. Ni, Z. A. Ren, G. C. Che, H. P. Yang, Z. Y. Liu, Z. X. Zhao, Chin. Phys. Lett. {\bf 18}, 816(2001).
\end{references}
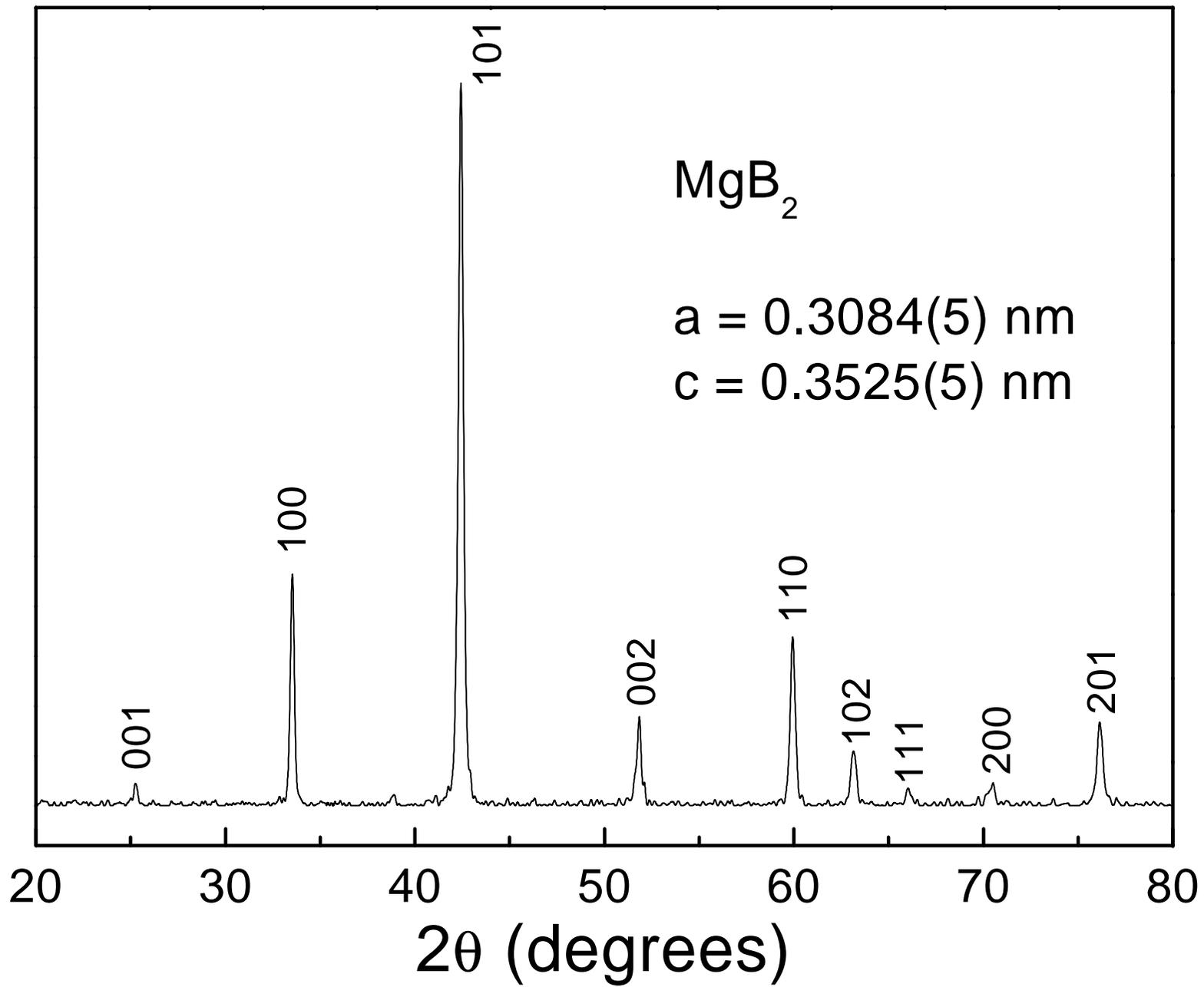
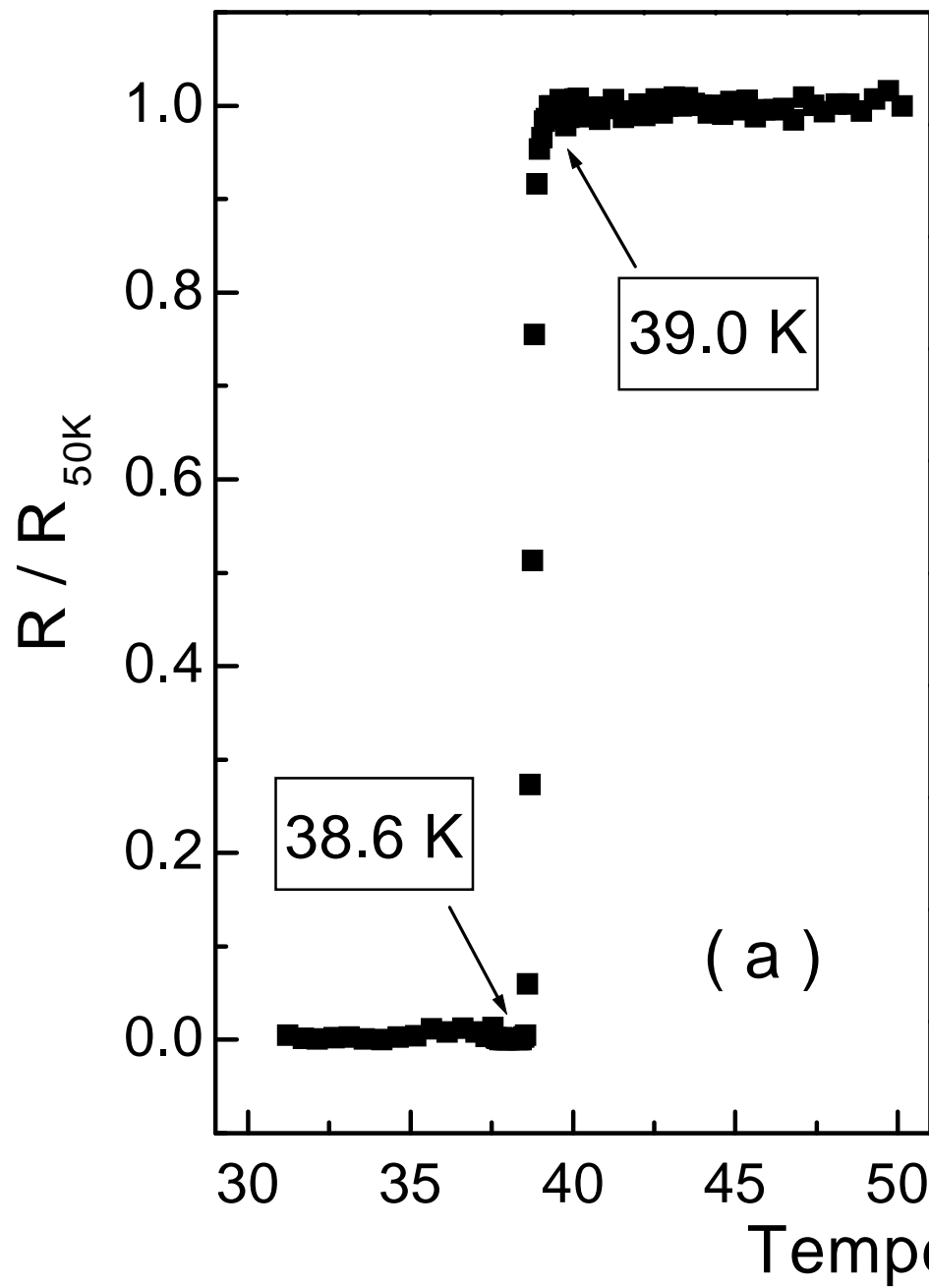
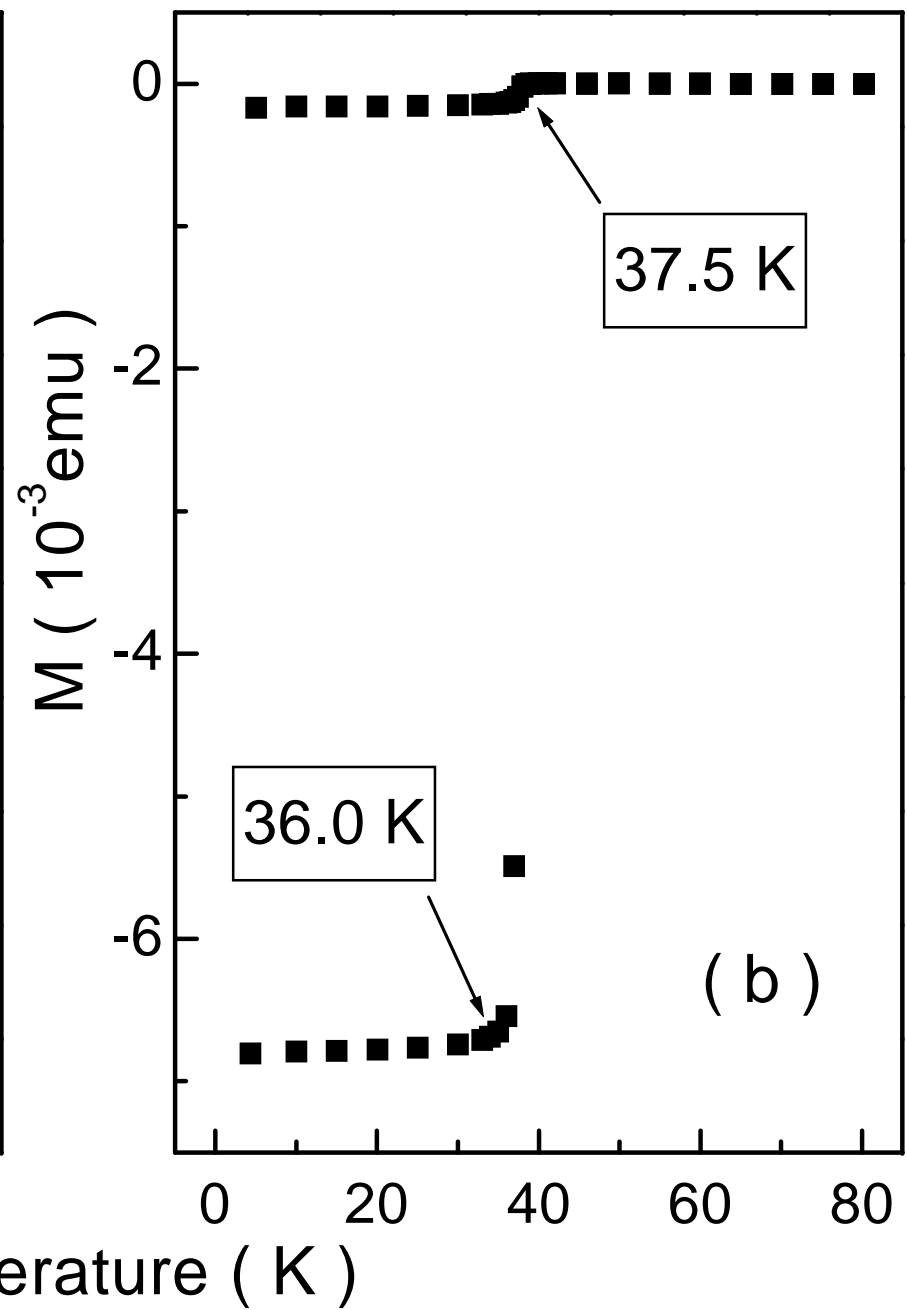
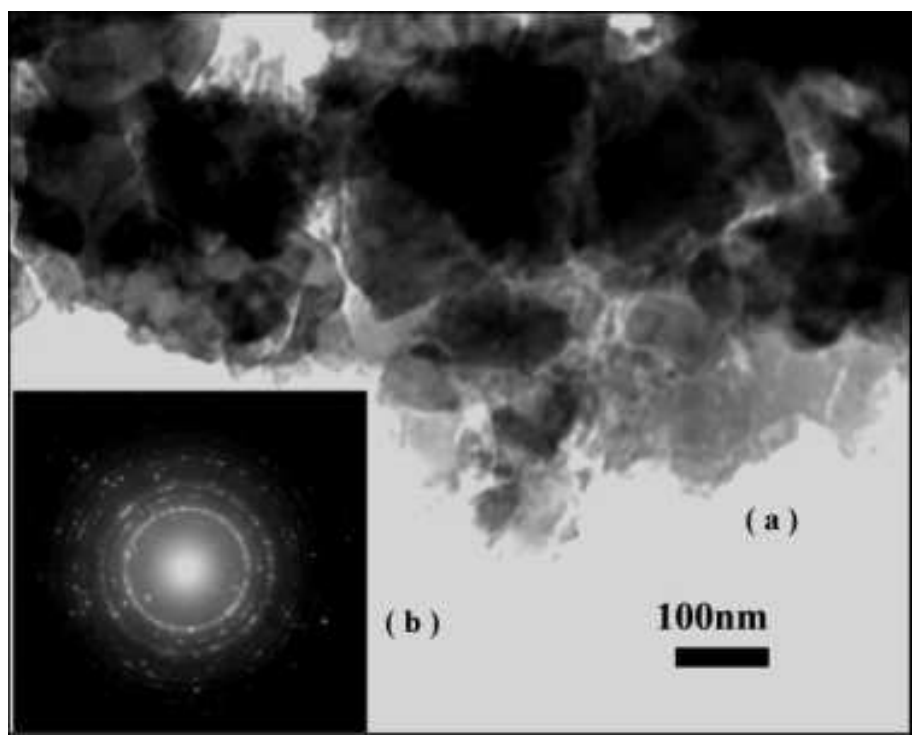
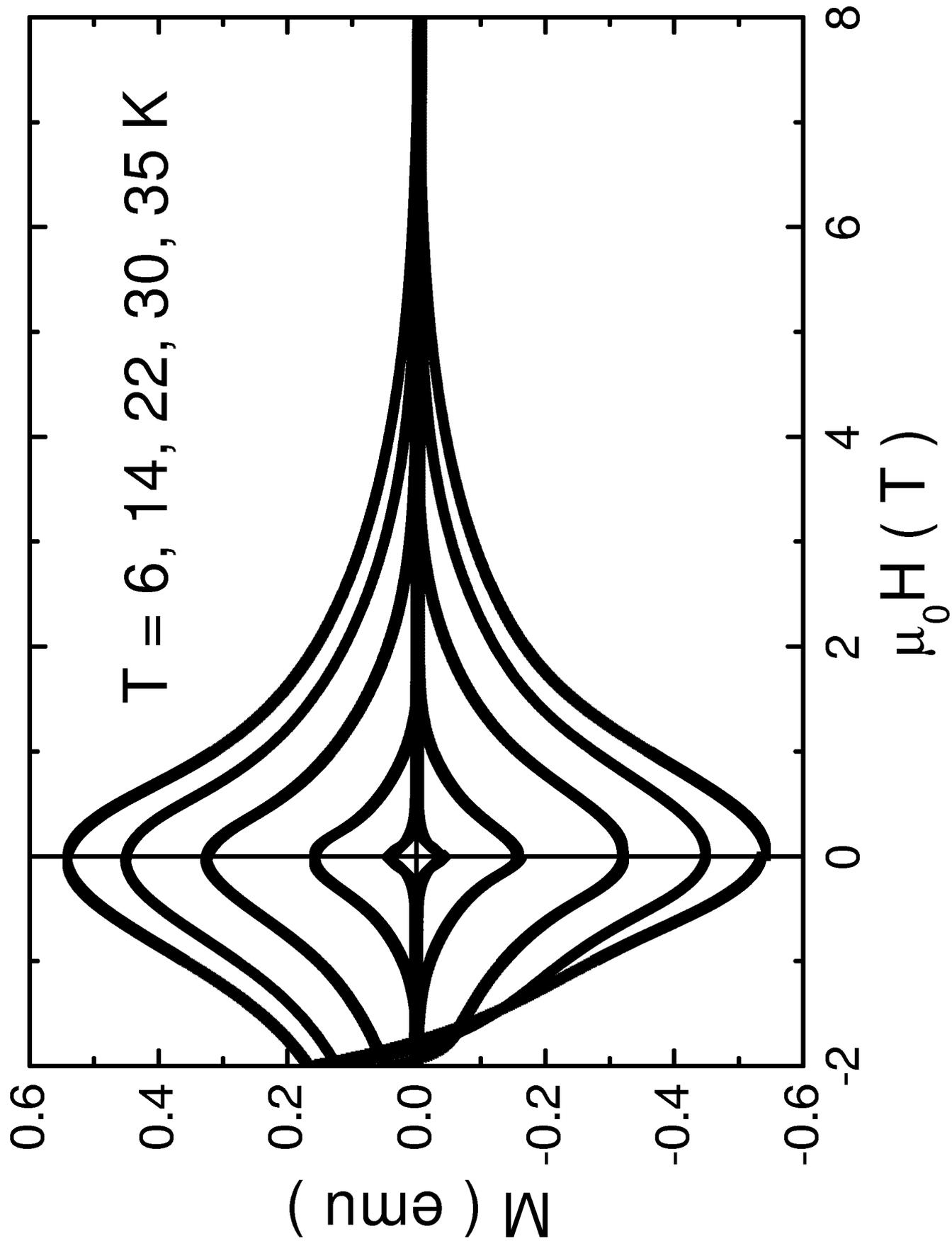
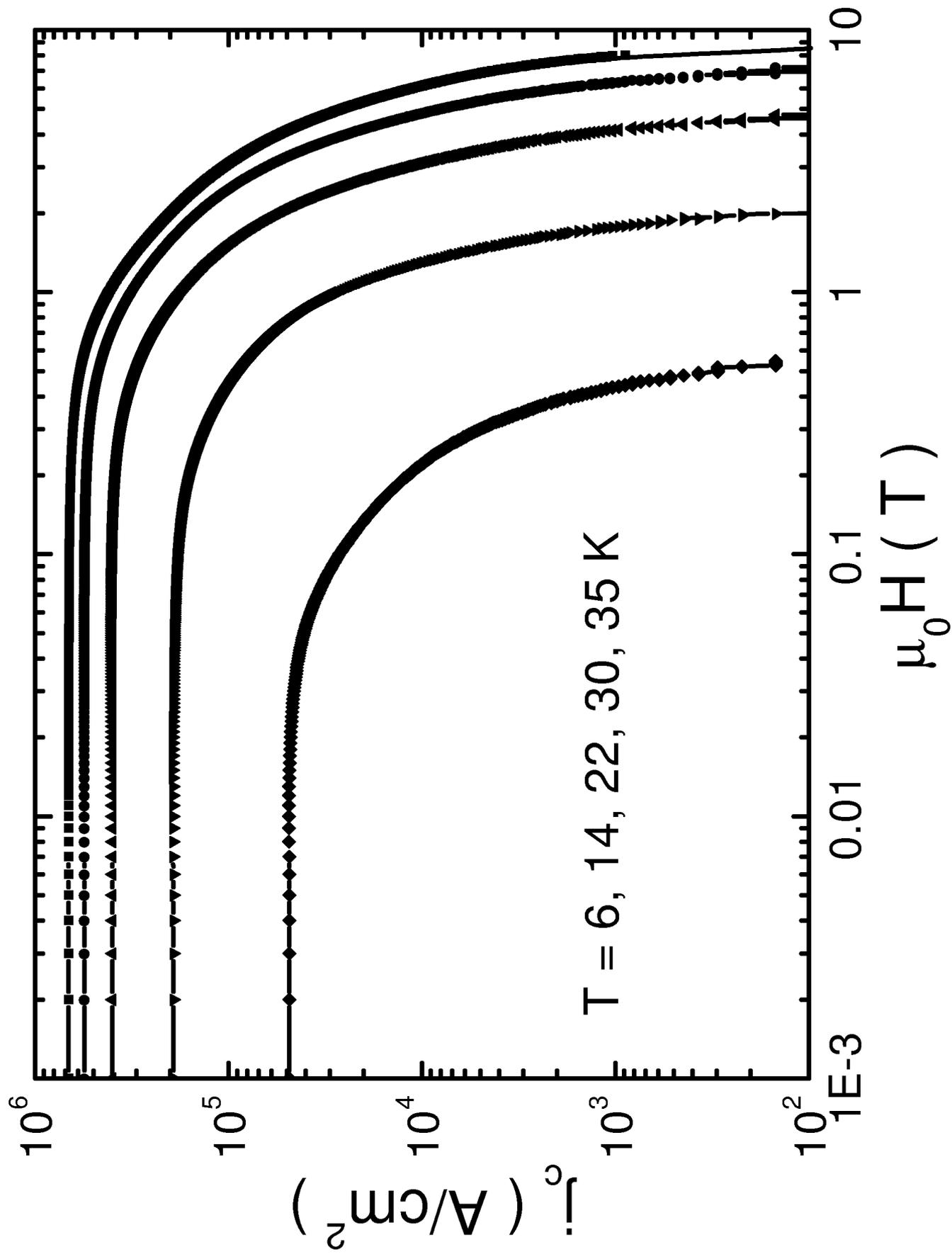
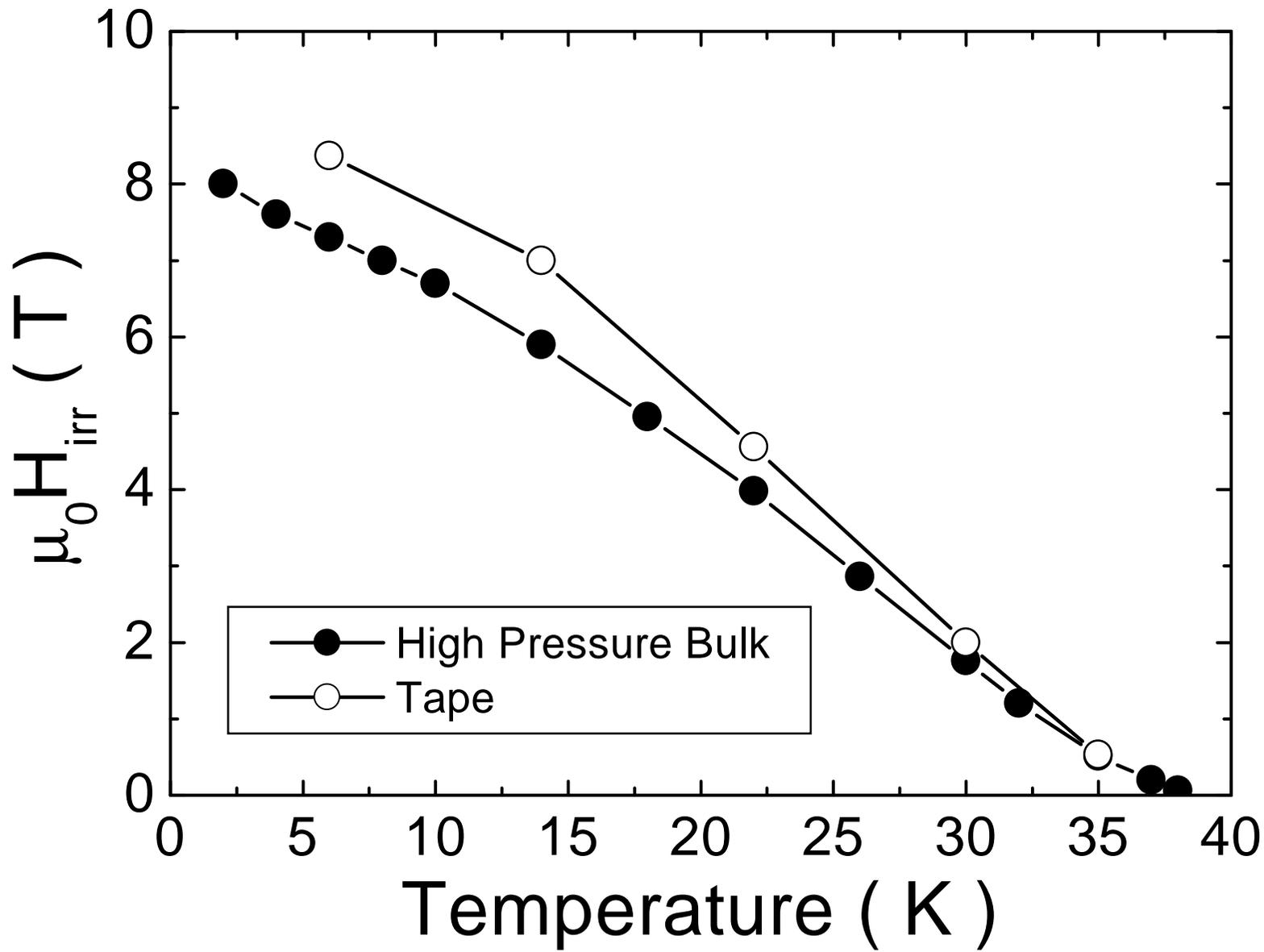
\end{document}